\documentclass[%
reprint,
superscriptaddress,
 amsmath,amssymb,
 aps,
prb,
floatfix,
]{revtex4-2}

\usepackage{mathtools}
\usepackage{comment}
\usepackage{appendix}
\usepackage{lipsum}
\usepackage{xcolor}
\usepackage{physics}
\usepackage{bbm}
\usepackage{graphicx}% Include figure files
\usepackage{dcolumn}% Align table columns on decimal point
\usepackage{bm}% bold math
\begin{document}

\title{Tuning terahertz transitions in cyclo[$n$]carbon rings}

\author{R. A. Ng}
\affiliation{Physics Department, De La Salle University, 2401 Taft Avenue, 0922 Manila, Philippines}

\author{M. E. Portnoi}
\affiliation{Physics and Astronomy, University of Exeter, Stocker Road, Exeter EX4 4QL, United Kingdom}
%\affiliation{ITMO University, St. Petersburg 197101, Russia}

\author{R. R. Hartmann}
\email{richard.hartmann@dlsu.edu.ph}
\affiliation{Physics Department, De La Salle University, 2401 Taft Avenue, 0922 Manila, Philippines}

\begin{abstract}
We develop an analytic model for an ideal polyyne ring which describes the induced THz gap in the molecular spectrum due to the Stark effect. This simple model can also be used to describe an odd-dimered cyclocarbon which has undergone a spontaneous symmetry-breaking event (due to the Jahn-Teller effect) as an effective dipole across an ideal ring. We show that both the size of the gap, and the strength of optical transitions across it, can be modulated by varying the external electric field strength. A THz emission scheme based on optical excitation is proposed, thus paving the way for a new class of THz emitters based on cyclo[$n$]carbon.
\end{abstract}

\maketitle

%\section*{Introduction}
The terahertz (THz) gap is a region in the electromagnetic spectrum sandwiched between the microwave and infrared regimes which suffers from a lack of portable, affordable, coherent sources and detectors that operate at room temperatures~\cite{dragoman2004terahertz, lee2007searching}. One pathway to bridge the THz gap is to utilize low-dimensional forms of carbon~\cite{hartmann2014terahertz} such as graphene, carbon nanotubes%~\cite{kibis2007generation, portnoi2008terahertz, hartmann2019interband}
, and stable linear carbon chains that were recently synthesized via laser ablation~\cite{pan2015carbyne, kutrovskaya2020excitonic}. We propose a new avenue of exploration for THz generation, utilizing monocyclic rings of carbon, subject to an experimentally attainable electric field applied in the plane of the ring.
%We propose a new avenue of exploration for THz generation, utilizing monocyclic rings of carbon subjected to an applied electric field. 

In contrast to the conjugated $\pi$-bond systems of fullerenes, carbon nanotubes, and graphene which have a coordination number of three, cyclo[$n$]carbon ($\mathrm{C}_n$), a ring of $n$ carbon atoms, has a coordination number of two. Hence there are two possible bonding configurations for cyclocarbons: polyynic, i.e., alternating single and triple bonds ($D_{nh/2}$ symmetry), or cumulenic, i.e., consecutive double bonds ($D_{nh}$ symmetry). Although evidence for the existence of cyclocarbon in the gaseous phase has been around since the 1980s~\cite{mcelvany1986ion, mcelvany1987ion, mcelvany1988reactions, parent1989investigations, yang1988ups, von1991structures, diederich1989all}, the high reactivity of $\mathrm{C}_n$ made its isolation in the condensed phase difficult to realize. A comprehensive review of the history of carbon rings can be found in Ref.~\cite{anderson2021short}. It was not until the pioneering work of Ref.~\cite{kaiser2019sp} that $\mathrm{C}_{18}$ was successfully isolated and its structure characterized via atomic force microscopy to show a polyynic structure.

%There are two possible bonding configurations for cyclocarbons: polyynic, i.e., alternating single and triple bonds ($D_{9h}$ symmetry), or cumulenic, i.e., consecutive double bonds ($D_{18h}$ symmetry). 

Symmetry plays a vital role in determining the electronic and optical properties of polymer molecules, and the selection rules of various molecules can be obtained within the group theory formalism~\cite{bozovic1978irreducible, bozovic1981irreducible, damnjanovic1983selection, damnjanovic1983standard, damnjanovic1984selection}. In this paper, we shall determine the optical selection rules via a finite-matrix tight-binding formalism, which also allows for a quantitative description of the system.
There have been various theoretical studies of cyclocarbons focusing on their structure, relative energetics, vibrational frequencies, and spectra~\cite{yang1988ups,martin1996structure,belau2007ionization,fowler2009double,brito2018quantum,baryshnikov2019cyclo,liu2020sp,liu2020sp2,anderson2021short}. Notably, odd- and even-dimered rings have been shown to alternate in their relative stability~\cite{hutter1994structures,yen2015use}.
The interplay between H\"{u}ckel's rule and the Jahn-Teller effect plays an important role in determining the ground-state configuration of cyclo[$n$]carbon. For $\mathrm{C}_{4p}$ rings, where $p$ is an integer, the first-order Jahn-Teller effect leads to single/triple bond alternation~\cite{bylaska1998development, bylaska2000small}; whereas for $\mathrm{C}_{4p+2}$ rings, bond length alternation is caused by the second-order Jahn-Teller effect~\cite{hong2020competition}. The Jahn-Teller effect can also give rise to bond angle alternation in both cumulenic and polyynic rings \cite{scriven2020molecular, stanger2006nucleus}, yielding four possible symmetries. There has been a great deal of controversy regarding the exact form of the ground state structure of various freestanding cyclo[$n$]carbons, with different computational models yielding drastically different results~\cite{plattner1995c18, seenithurai2020tao}. However, in the case of the recently isolated $\mathrm{C}_{18}$, placed on top of an inert substrate, the structure was experimentally shown to be polyynic~\cite{kaiser2019sp,scriven2020synthesis3}.

%There are a number of proposals to use the lifting of degeneracy in carbon-based nanostructures for THz applications including:

There are a number of proposals to use the lifting of symmetry-induced degeneracy of energy levels in carbon-based nanostructures for THz applications, including: magnetic field~\cite{portnoi2008terahertz} and strain-induced~\cite{hartmann2015terahertz} gap opening in carbon nanotubes, edge effects in graphene nanoribbons~\cite{hartmann2019interband} and finite polyyne chains~\cite{hartmann2021thz}, as well as tunable bipolar graphene waveguides~\cite{hartmann2020bipolar} and double-well systems~\cite{hartmann2020guided}. There are also several proposals for generating THz radiation which are based on the lifting of degeneracy between energy levels in semiconducting quantum rings that utilize a combination of magnetic and in-plane electric fields~\cite{alexeev2012electric, alexeev2012terahertz, alexeev2013aharonov}, as well as inducing a double quantum well via two electrostatic lateral gates~\cite{collier2017tuning}. 
%~\cite{collier2017tuning, collier2019terahertz}. 
%There are several proposals for generating THz radiation which are based on the lifting of degeneracy between energy levels in semiconducting quantum rings which utilize a combination of magnetic and in-plane electric fields~\cite{alexeev2012electric, alexeev2012terahertz, alexeev2013aharonov}, as well as inducing a double quantum well via two electrostatic lateral gates~\cite{collier2017tuning, collier2019terahertz}. 
%Degenerate energy levels in double-well systems \cite{hu1991feasibility, hartmann2020guided} and bipolar waveguides~\cite{hartmann2020bipolar} are also promising candidates as the building blocks of THz devices. 
%cannot be lifted in the first order due to the inversion symmtry. in the absences of a magnetic field...lifitng is not possible in the first in magnetic field.. we need to mention inversion, first order is possible in drastic difference from the even number of dimers, which are more simliar to continous rings due to inversion symmetry, symmtric has inversion centre, odd has reflection but not inversion
In this paper we show that the absence of inversion symmetry results in a striking contrast between odd- and even-dimered polyynic rings. Namely, for a ring with an odd number of dimers, the lifting of the degeneracy is linear in the magnitude of the applied electric field, while in a ring composed of an even number of dimers the presence of an inversion center forbids linear-in-electric-field level splitting, very much like the semiconducting quantum rings mentioned above.
%is very much like the semiconducting quantum rings mentioned above, where the presence of an inversion center forbids the linear in electric field splitting of degenerate levels.
%means the degeneracy can only be lifted for higher orders of perturbation theory.
%As we show in this paper, the absence of inversion symmetry results in a striking contrast between a polyynic ring with an odd and even number of dimers: Namely, for an odd dimer ring, the lifting of the degeneracy can be achieved via a first order perturbation of an external field, while for a ring possessing an even number of dimers it cannot. Indeed, a ring formed of an even number of dimers are very much like the semi-conducting quantum rings mentioned above, where the presence of an inversion centre means the degeneracy can only be lifted in the second order.}
%\textbf{Due to the symmetry of the system, an ideal polyynic ring possesses degenerate energy levels, which is in stark contrast to a semiconducting quantum ring that requires a magnetic field to create a degenerate level \cite{alexeev2012electric}.
Nonetheless, in a similar manner to semiconducting quantum rings, a perturbation can be used to split the degenerate levels of cyclocarbon and, due to symmetry, the two emergent levels can support optical transitions between them. This splitting can be calculated using time-independent degenerate perturbation theory. However, it should be noted that spontaneous symmetry-breaking can occur due to the second-order Jahn-Teller effect~\cite{pereira2020spontaneous}. In this instance, the degeneracy is removed, and there is an intrinsic gap associated with the originally degenerate levels. When spontaneous symmetry-breaking occurs, an effective dipole is set up across the molecule. This system can be modelled as an ideal polyynic ring subject to an effective electric field, where the existing field can be tuned by an external electric field to fall within the THz range. Therefore, cyclo[$n$]carbon is a potential candidate for THz applications. It should also be noted that computational models which take into account the Jahn–Teller effect in $\mathrm{C}_{18}$ show that an applied electric field produces a strong regulation effect on the ring's geometry and electronic structure~\cite{lu2021ultrastrong}. Indeed, for experimentally attainable fields the size of the gap between the highest occupied molecular orbital (HOMO) and the lowest unoccupied molecular orbital (LUMO) can be reduced from the UV into the visible frequency range, thus making any optical pumping scheme more accessible.
%Indeed, for experimentally attainable fields the size of the HOMO-LUMO gap can be reduced from the UV into the visible frequency range thus making any optical pumping scheme more accessible.

For an ideal polyynic ring, in the absence of an applied electric field, the eigenvalues and eigenfunctions can be solved exactly. For a ring composed of an even number of dimers, both the highest energy level and the lowest energy level are non-degenerate, together with the HOMO and LUMO: all the other energy levels are twofold-degenerate. In contrast, if there is an odd number of dimers, the highest and lowest energy levels alone are non-degenerate, while the rest are twofold-degenerate.

In what follows, for an ideal polyynic ring composed of an odd number of dimers, we use first-order degenerate perturbation theory to determine the energy level splitting and optical selection rules, supported by full matrix calculations.
\begin{figure}
    \centering
    \includegraphics[width=0.5\textwidth]{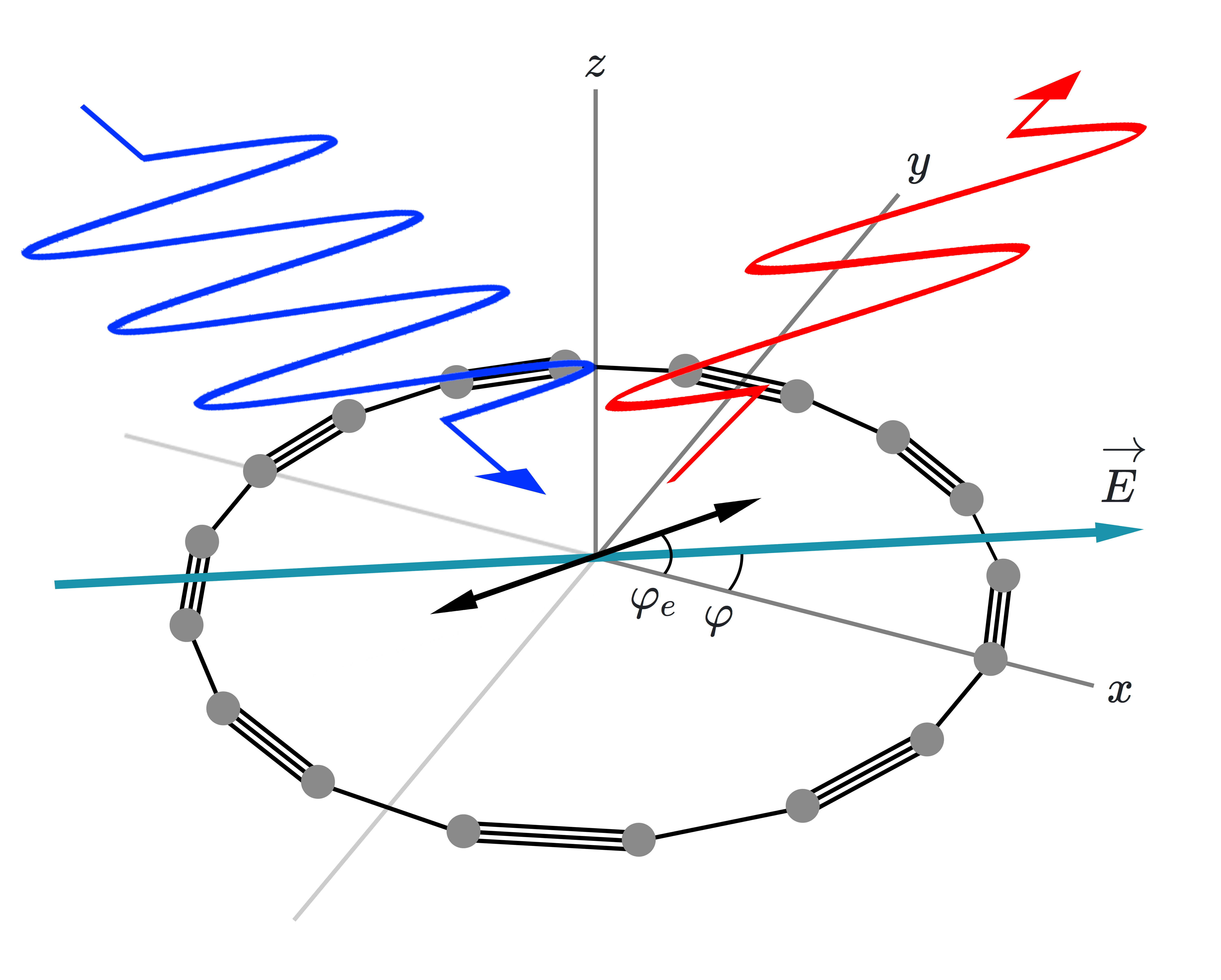}
    \caption{A schematic representation of cyclo[$n$]carbon with a polyynic structure (i.e., alternating single and triple bonds) composed of an odd number of dimers. The ring is subjected to an in-plane electric field $\boldsymbol{E}$, depicted by the green arrow, and oriented at an angle $\varphi$ from the $x$-axis. The black double arrow is the projection of the incident, linearly polarized light (blue wave) onto the plane of the ring, where $\varphi_e$ is the angle between this projection and the horizontal axis. A sketch of the emitted THz photon is shown by the red wave.
    }
    \label{fig:schematic}
\end{figure}

%Although the exact symmetry of cyclo[18]carbon is a subject of debate, the emerging consensus it is polyynic, i.e., a ring of alternating single and triple bonds, rather than cumulenic, i.e., consecutive double bonds.
%SAMPLE LINES: carbon nanotube can be considered as a rolled graphene sheet. Similarly, linking both ends of a carbon chain can form a cyclocarbon.
%Four possible symmetries: cumulenic with and without bond angle alternation, polyynic with and without bond angle alternation. $D_\mathrm{Nh}$, $C_\mathrm{N/2h}$, 

%\section*{Theory}
We shall now present a simple analytic model to describe the optical transitions between the electronic states of an ideal polyynic ring. We adopt the same nearest-neighbor tight-binding formalism which is commonly used in carbon-based materials \cite{dresselhaus1998physical}. This model yielded spectacular success in describing both their electronic and optical properties, especially in the case of carbon nanotubes and graphene~\cite{neto2009electronic,hartmann2014terahertz}.
%\cite{hartmann2014terahertz, hartmann2015terahertz, portnoi2015terahertz, hartmann2016exciton, hartmann2017pair, hartmann2019interband}.
Using the nearest-neighbor tight-binding model~\cite{dresselhaus1998physical}, the Hamiltonian of an ideal polyynic ring composed of $n$ carbon atoms (where $n$ is an even integer), schematically depicted in Fig.~\ref{fig:schematic},  can be written as:
\begin{equation}
H_{ij}=\left(\begin{array}{cccccc}
0 & b_{1} &  &  &  & b_{n}\\
b_{1} & \ddots & \ddots\\
 & \ddots & \ddots & b_{j}\\
 &  & b_{j} & \ddots & \ddots\\
 &  &  & \ddots & \ddots & b_{n-1}\\
b_{n} &  &  &  & b_{n-1} & 0
\end{array}\right),
\label{eq:matrix}
\end{equation}
where $b_j=t$~($T$) when $j$ is odd (even). It should be noted that as the number of atoms approaches infinity, the band structure transforms from discrete energy levels to the continuous dispersion of an infinite polyyne chain. Therefore, we shall use the same tight-binding parameters associated with the infinite chain; namely, $t=-3.548$~eV and $T=-4.657$~eV~\cite{al2014electronic}. The eigenvalues and eigenvectors of the Hamiltonian given in  Eq.~(\ref{eq:matrix}) can be solved exactly. The energy levels of a polyynic ring composed of $n$ atoms is given by the expression 
\begin{equation}
\varepsilon_{l}=s_{1}\sqrt{\left(T-t\right)^{2}+4tT\cos^{2}\left(\frac{2\pi l}{n}\right)},
\end{equation}
where $l=1,\,2,\,\ldots n/2$ and $s_{1}$ can take the values of $1$ and $-1$. The $j^\mathrm{th}$ component of the corresponding eigenfunctions,
$\Psi_{l,\,s_{1}}=\left(\psi_{1},\psi_{2},\ldots,\psi_{n}\right)^{\mathrm{T}}$ , can be written as
\begin{equation}
\psi_{j}=\frac{e^{ik_{l}aj}}{\sqrt{n}}\begin{cases}
1 & \mathrm{for}\,j=\mathrm{odd}\\
s_{1}\frac{f_{l}}{\left|f_{l}\right|} & \mathrm{for}\,j=\mathrm{even}
\end{cases},
\end{equation}
where $f_{l}=te^{-ik_{l}a}+Te^{ik_{l}a}$ and $k_{l}=2\pi l/\left(na\right).$ For a ring consisting of an odd number of dimers, all the energy levels are twofold-degenerate except for the $l=n/2$ energy levels, whereas for the case of an even number of dimers, all the levels are degenerate except for the $n/2$ and $n/4$ levels. 

For an ideal ring, applying an in-plane electric field, $\boldsymbol{E}$, lifts the degeneracy of the system. In the presence of an in-plane electric field the perturbation to the Hamiltonian, $\delta H_{ij}$, is given by the expression
\begin{equation}
\delta H_{ij}=V_{0}\cos\left(\frac{2\pi j}{n}-\varphi\right)\delta_{ij},
\end{equation}
where $V_{0}=eER$, $R$ is the radius of the carbon ring, $e$ the elementary charge and $\varphi$ the angle of the electric field relative to the $x$-axis (the geometry of the problem is shown in Fig.~\ref{fig:schematic}). Let us now consider two degenerate states, $\Psi_{l,\,s_{1}}$ and $\Psi_{l',\,s_{2}}$, belonging to a ring composed of an odd number of dimers. Using degenerate first-order perturbation theory, the first-order correction term to the HOMO/LUMO level is
$\delta \varepsilon=s_{3}\alpha_{n}\left|V_{0}\right|/2$,
where $s_{3}=1,-1$ and $\alpha_n=\Delta_{\infty}\cos\left(\pi/n\right)/\Delta_{n}$ is a geometric parameter defined by: the HOMO-LUMO gap of a linear polyynic chain, $\Delta_{\infty}=2\left|T-t\right|$, and the HOMO-LUMO gap of the polyynic ring itself, $\Delta_{n}=2\left|\varepsilon_{l}\right|$, where $l=(n-2)/4$. Therefore, the size of the gap between the HOMO (LUMO) and HOMO$-1$ (LUMO$+1$) is
%Using degenerate first-order perturbation theory, the first-order correction term to the HOMO/LUMO level is $\delta E=s_{3}\alpha_{n}\left|V_{0}\right|/2$, where $s_{3}=1,-1$, $\alpha_n=\Delta_{\infty}\cos\left(\pi/n\right)/\Delta_{n}$ is a geometric parameter, defined by the HOMO-LUMO gap of a linear polyynic chain, $\Delta_{\infty}=2\left|T-t\right|$, and the HOMO-LUMO gap of the polyynic ring itself, $\Delta_{n}=2\left|\varepsilon_{l}\right|$, where $l=(n-2)/4$. Therefore, the size of the HOMO-HOMO$-1$ gap is
\begin{equation}
E_g= \alpha_{n}\left|V_{0}\right|.
\label{eq:split}
\end{equation} 
Hence, a 1 THz gap can be opened by applying an electric field of the order of $10^{7}~\mathrm{V/m}$, which is easily achievable if the rings are encapsulated in a material with high dielectric strength, e.g., h-BN. It should be noted that so far we have assumed an equal distance between atoms, but the effect of bond length alternation can be readily accommodated in our model. However, it only results in a small correction factor to the coefficient $\alpha_n$ appearing in Eq.~(\ref{eq:split}), which should be changed to $\widetilde{\alpha}_n\approx\alpha_{n}[1+\widetilde{\delta}\tan\left(\pi/n\right)]$, where $\widetilde{\delta}=2\pi\delta_{c}/\left(cn\right)$, and $\delta_{c}$ is the difference in length between a single and triple bond, and $c$ is the lattice constant.
%Since this correction factor is negligibly small, and does not qualitatively change the results.

\begin{figure}
    \centering
    \includegraphics[width=0.45\textwidth]{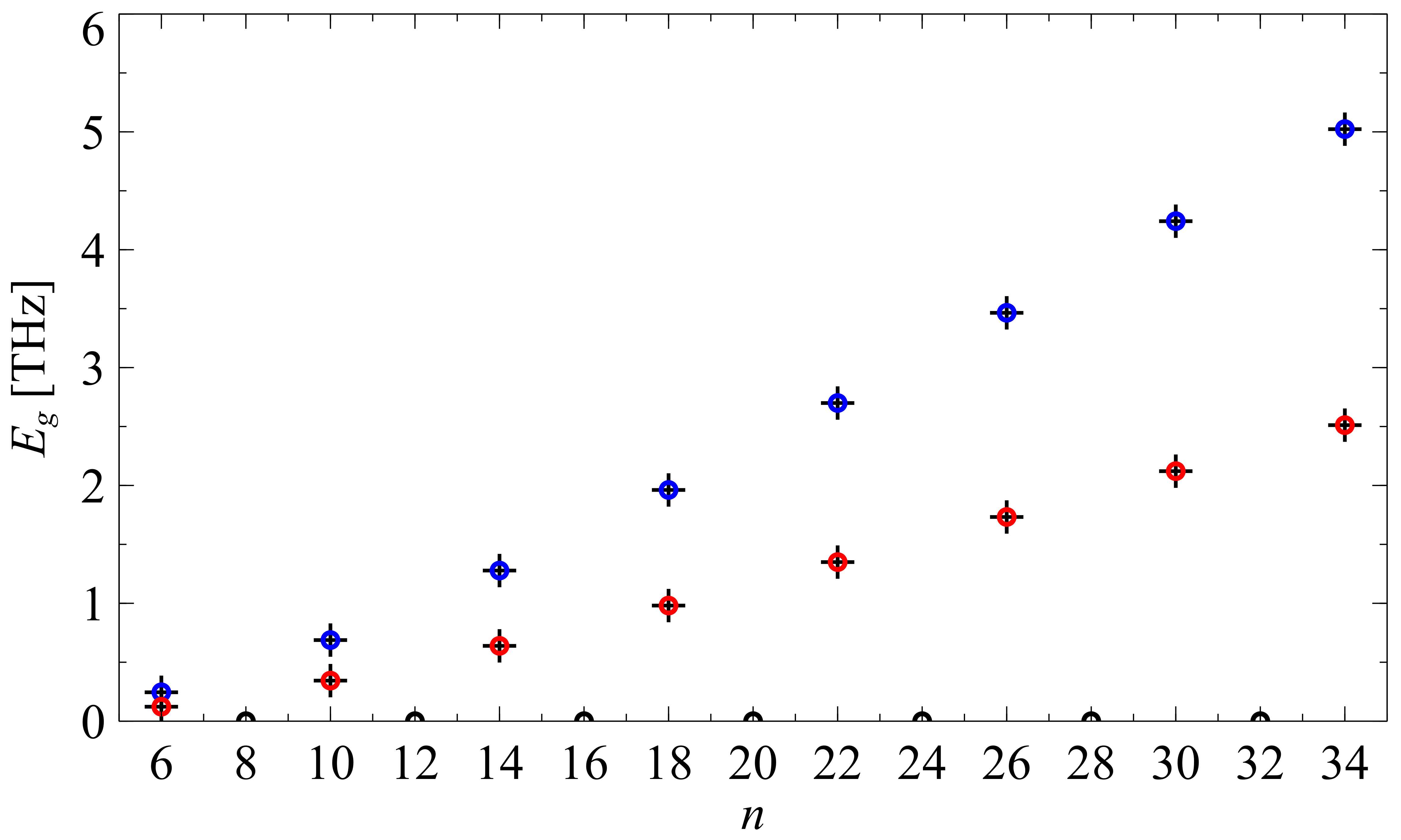}
    \caption{
    The HOMO-HOMO$-1$ gap of cyclo[$n$]carbon for the case of an in-plane electric field strength of $1.8$ and $3.6\times10^{7}~\mathrm{V/m}$. The numerical solutions of the nearest-neighbor tight-binding matrix are depicted by the black crosses, whereas the analytic solutions are depicted by the red and blue circles which correspond to the field strengths of $1.8$ and $3.6~\times10^{7}~\mathrm{V/m}$, respectively. For contrast, the black circles show the energy level splitting for even-dimered rings in the higher electric field. The analytic expression, valid only for an odd-dimered ring, is given by $E_g=\alpha_{n}\left|eER\right|$, where $e$ is the elementary electric charge, $E$ the electric field strength, and $R$ the radius of the ring. The geometric parameter $\alpha_n=\Delta_{\infty}\cos\left(\pi/n\right)/\Delta_{n}$ is defined by the HOMO-LUMO gap of a linear polyynic chain, $\Delta_{\infty}=2\left|T-t\right|$, and the HOMO-LUMO gap of the ring itself, $\Delta_{n}=2[\left(T-t\right)^{2}+4tT\cos^{2}\left(2\pi l/n\right)]^{1/2}$, with $l=(n-2)/4$. The tight-binding parameters used are $t=-3.548$~eV and $T=-4.657$~eV, respectively.
    %$\Delta_{n}=2[\left(T-t\right)^{2}+4tT\cos^{2}\left(2\pi l/n\right)]^{1/2}$ 
}
    \label{fig:E_gap}
\end{figure}

For an ideal ring composed of an even number of dimers, the first-order correction term is zero; therefore, higher-order perturbation terms need to be considered in order to calculate the true value of the energy level splitting. However, for experimentally attainable electric fields, $V_{0}$ is of the order of THz; therefore any energy gap opening associated with higher-order perturbation terms is beyond the spectral range of interest. This is also true for the splitting of other higher/lower levels. In Fig.~\ref{fig:E_gap}, we plot the numerical solutions for the tight-binding Hamiltonian, given in Eq.~(\ref{eq:matrix}), together with the analytical expression given in Eq.~(\ref{eq:split}), for two different electric field strengths, $1.8$ and $3.6\times10^{7}~\mathrm{V/m}$, respectively. Regardless of ring size, any desired THz gap can be achieved by simply modulating the strength of the electric field. In the same figure we plot the result of the numerical calculation of the energy level splitting, $E_g$, for rings composed of an even number of dimers, which is indeed negligibly small. It resembles the electric-field-induced energy level splitting of double-degenerate levels in the continuum model for semiconducting rings~\cite{fischer2009exciton,alexeev2012electric}, in which the linear-in-electric-field splitting requires a magnetic flux equal to half of a magnetic flux quantum through the ring. Thus, adding an extra dimer to a $\mathrm{C}_{16}$ ring results in linear-in-electric-field level splitting which is equivalent to using a magnetic field of the order of $10^4$~T for the same ring.

As can be seen from Eq.~(\ref{eq:split}), the magnitude of the splitting of the two degenerate states comprising the HOMO level of an ideal odd-dimered ring is linearly proportional to the applied electric field and the radius of the ring. An in-plane electric field can be generated by applying a voltage across two lateral gates, situated on either side of the ring. For the case of a $\mathrm{C}_{18}$ molecule, an applied electric field strength of $1.8\times10^{7}~\mathrm{V/m}$ yields a gap of approximately $1$~THz. However, as mentioned in the introduction, it has been predicted that $\mathrm{C}_{18}$ may exhibit spontaneous symmetry-breaking, which lifts the degeneracy of the energy levels. In this instance, the predicted energy gap associated with the splitting of the HOMO level is $0.140$~eV~\cite{pereira2020spontaneous}. But it is well known that such calculations overpredict the value of inter-level spacings, and must be scaled appropriately. One scaling factor which may be implemented is $\mathrm{C}_{60}^{\mathrm{Exp}}/\mathrm{C}_{60}^{\mathrm{Theo}}$~\cite{lof1992band}, where $\mathrm{C}_{60}^{\mathrm{Exp}}$ is the HOMO-LUMO gap of fullerene $\mathrm{C}_{60}$ in experiment (2.30 eV), and $\mathrm{C}_{60}^{\mathrm{Theo}}$ is that which was obtained computationally. Using this scaling factor, the experimental HOMO-LUMO gap of $\mathrm{C}_{18}$ has been predicted to be 2.72~eV~\cite{li2020potential}. Implementing the aforementioned scaling factor, the predicted gap between the HOMO and HOMO$-1$ levels reduces to $\approx0.071$~eV, which corresponds to 17~THz. Thus, a polyynic ring which has undergone a spontaneous symmetry-breaking event can be modelled as an ideal ring subject to an effective electric field, i.e., the second-order Jahn-Teller effect induces an effective dipole across the ring. Due to spontaneous symmetry-breaking, $\mathrm{C}_{18}$ inherently possesses a gap which corresponds to frequencies within the upper THz regime. However, this gap can still be modified by the application of an external electric field, and is therefore highly tunable. Indeed, applying an electric field of the correct magnitude along the right direction could suppress the inherent gap altogether. However, to determine the exact size and behavior of the gap in an applied electric field requires state-of-the-art quantum chemistry calculations.

%Indeed, it can be seen from the distribution of the molecular orbitals that there is an effective dipole across the ring~\cite{pereira2020spontaneous} 

Within the framework of degenerate first-order perturbation theory, the splitting of the HOMO and LUMO states can be analyzed independently of the other levels. The new eigenstates of the doublet emerging from the application of an electric field can be written as a linear combination of the unperturbed eigenstates, $\left|\Psi_{m}\right\rangle$:
\begin{equation}
\left|\Psi_{\pm}\right\rangle =\frac{1}{\sqrt{2}}\left[\left|\Psi_{l}\right\rangle \pm\frac{\Delta}{\left|\delta \varepsilon\right|}\left|\Psi_{l+1}\right\rangle \right],
\end{equation}
where $\Delta=V_{0}\left[1+\left(f_{l+1}/f_{l}\right)\right]e^{i\varphi}/4$, $l=\left(n-2\right)/4$, $\left|\Psi_{+}\right\rangle$  is the eigenstate of the highest energy level of the doublet, and $\left|\Psi_{-}\right\rangle$  the lowest. The strength of the dipole matrix element of transition between them, caused by linearly polarized light, with the polarization vector $\hat{\boldsymbol{e}}=\left(\cos\varphi_{e},\,\sin\varphi_{e}\right)$ is
\begin{equation}
\left|\left\langle \Psi_{+}\left|\hat{\boldsymbol{e}}\cdot\boldsymbol{r}\right|\Psi_{-}\right\rangle \right|=\frac{\alpha_{n}R}{2}\left|\sin\left(\varphi_{e}-\varphi\right)\right|.
\end{equation}
According to Fermi's golden rule, the transition probability per unit time between the higher and lower states of the doublet is proportional to the absolute value of the velocity matrix element squared. The transition dipole moment between the two eigenstates can be written in terms of the velocity operator, using the relationship 
$\left|\left\langle \Psi_{+}\left|\hat{\boldsymbol{e}}\cdot\boldsymbol{r}\right|\Psi_{-}\right\rangle \right|=\left|\left\langle \Psi_{+}\left|\hat{\boldsymbol{e}}\cdot\hbar\boldsymbol{v}\right|\Psi_{-}\right\rangle \right|/\left|2\delta \varepsilon\right|$. Hence the absolute value of the matrix element of velocity is
\begin{equation}
\left|\left\langle \Psi_{+}\left|\hat{\boldsymbol{e}}\cdot\boldsymbol{v}\right|\Psi_{-}\right\rangle \right|=\frac{\alpha_{n}^{2}R}{2\hbar}\left|V_{0}\sin\left(\varphi_{e}-\varphi\right)\right|.
\label{eq:vme}
\end{equation}
%It should be noted that since $V_{0}=eER$, the velocity matrix element depends on the square of $R$. 
It can be seen from the above expression that for a given value of the electric field, the velocity matrix element depends on the square of the radius of the ring. %, product the square of the geometric factor $\alpha_n$. 
However, although the probability of the transition increases with increasing radius, so does the size of the gap between the HOMO and HOMO$-1$ levels. 

%Therefore, it should be noted that although the probability of the transition increases with increasing radius, so does the size of the gap between the HOMO and HOMO-1 levels. Equation~(\ref{eq:vme}) shows that the transition is also extremely anisotropic, and is maximal when the polarization vector is perpendicular to the electric field. %This highly polarization-sensitive transition can be exploited for THz applications, and will be discussed in the next section. %For linearly polarized light, the selection rules for transitions across the THz gap are highly anisotropic. This is in stark contrast to the case of circularly polarized light, where the dipole matrix element has no angular dependence. 

Equation~(\ref{eq:vme}) shows that the transition is also extremely anisotropic, and is maximal when the polarization vector is perpendicular to the electric field. Although we have discussed only transitions associated with linearly polarized light, transitions are also allowed for the case of circularly polarized light. Defining the polarization vector of circularly polarized light as $\hat{\boldsymbol{e}}_{\mathrm{cir}}=\left(1,\,\pm i \right)$, where the plus and minus signs correspond to left- and right-handed polarized light, respectively, the dipole matrix element is $\left|\left\langle \Psi_{+}\left|\hat{\boldsymbol{e}}_{\mathrm{cir}}\cdot\boldsymbol{r}\right|\Psi_{-}\right\rangle \right|=\alpha_{n}R/(2\sqrt{2})$. 

%This result is essentially a manifestation of the momentum alignment phenomenon in a two-dimensional (2D) Dirac material~\cite{saroka2018momentum, saroka2019interband}, where photo-excited carriers are created preferentially normal to the polarization vector for linearly polarized light, and isotropically for circularly polarized light. Since cyclocarbon can be considered as a linear polyyne chain linked together at both ends, the Hamiltonian given in Eq.~(\ref{eq:matrix}) can be reduced down to a $2\times2$ matrix, and for large enough $n$, near the band-edge has the same structure as a gapped 2D Dirac material.

\begin{figure}
    \centering
    \includegraphics[width=0.45\textwidth]{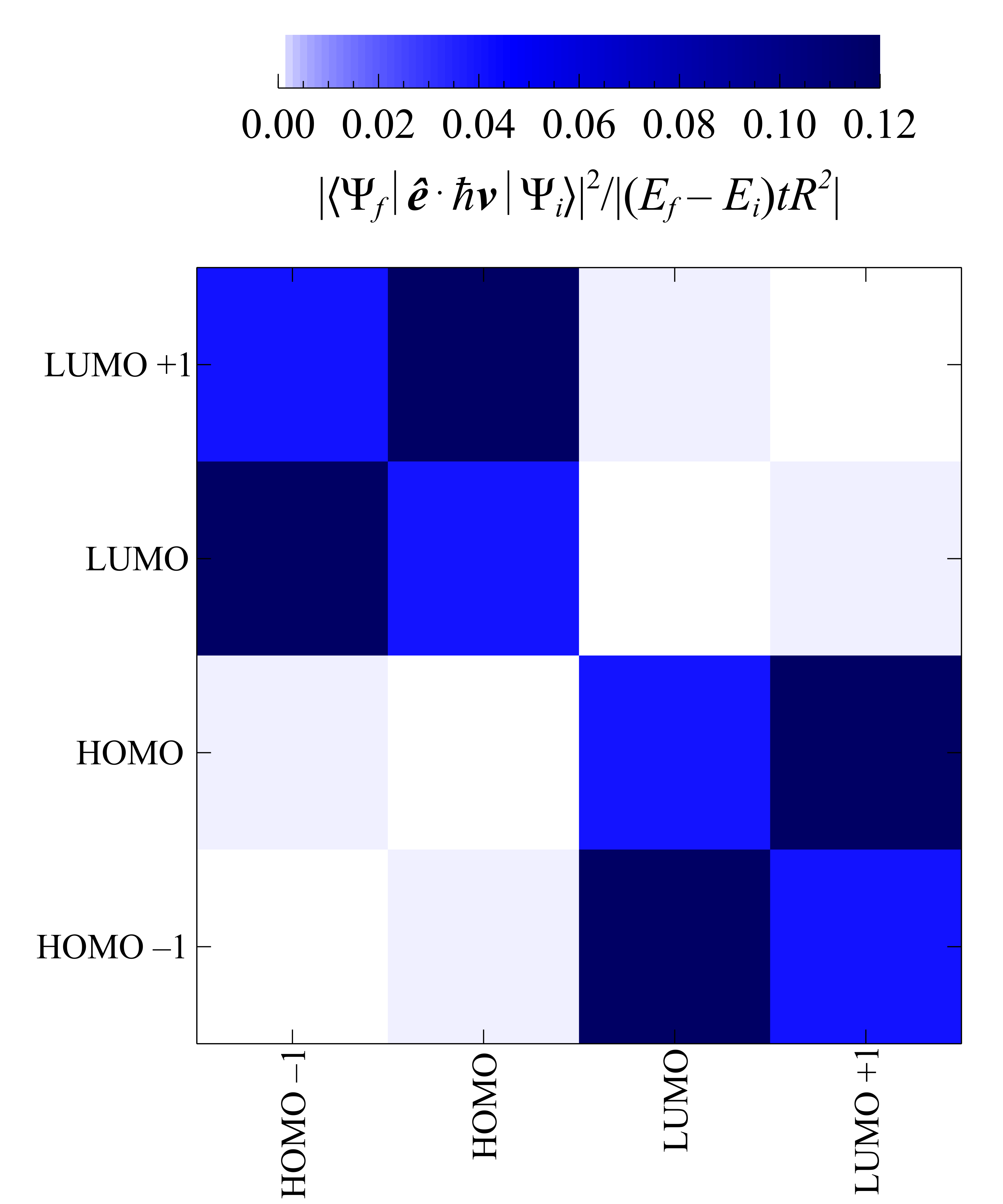}
    \caption{
The absolute value of the oscillator strength, normalized by a dimensionless constant, for transitions between energy
levels of a polyynic ring comprised of $n = 18$ atoms, subject to an electric field of $3.6~\times10^{7}~\mathrm{V/m}$. Here the polarization vector of the linearly polarized excitation is oriented at an angle of $\pi/3$ relative to the applied electric field, and the tight-binding parameters are $t=-3.548$ eV and $T=-4.657$~eV.}
    \label{fig:Oscillator_strength}
\end{figure}

Finally, it should be noted that since each set of doublet states associated with the HOMO and LUMO levels can be constructed as a linear combination of states with differing parity, all possible transitions between the four levels are %strongly 
allowed due to symmetry %(see Fig.~\ref{fig:THz_Scheme}).
(see Fig.~\ref{fig:Oscillator_strength}). Therefore this system supports both optical and THz transitions, which can be exploited for THz applications, and will be discussed shortly. %This will be discussed further in the next section. 
The velocity matrix element between the HOMO$-1$ (HOMO) and LUMO$+1$ (LUMO) is given by the expression
$\left|\left\langle \Psi'_{\mp}\left|\hat{\boldsymbol{e}}\cdot\hbar\boldsymbol{v}\right|\Psi_{\pm}\right\rangle \right|=\left|\beta_{n}\cos\left(\varphi_{e}-\varphi\right)\right|/\left[1\mp\left(E_{g}/\Delta_{n}\right)\right]$, while the velocity matrix element between the HOMO$-1$ (HOMO) and LUMO (LUMO$+1$) is given by 
$\left|\left\langle \Psi'_{\pm}\left|\hat{\boldsymbol{e}}\cdot\hbar\boldsymbol{v}\right|\Psi_{\pm}\right\rangle \right|=\left|\beta_{n}\sin\left(\varphi_{e}-\varphi\right)\right|$, where $\beta_{n}=\varepsilon_{0}R\sin\left(\pi/n\right)$, and $\Psi'_{\pm}$, $\Psi_{\pm}$ correspond to the HOMO and LUMO states, respectively. 
In Fig.~\ref{fig:Oscillator_strength}~ we plot $\left|\left\langle \Psi_{f}\left|\hat{\boldsymbol{e}}\cdot\hbar\boldsymbol{v}\right|\Psi_{i}\right\rangle \right|^{2}/\left|\left(E_{f}-E_{i}\right)tR^{2}\right|$ corresponding to the absolute value of the oscillator strength, normalized by a dimensionless constant $3\hbar^{2}/\left(2m_{e}tR^{2}\right)$ for the transitions between levels close to the HOMO and LUMO, for $n=18$, $\varphi-\varphi_e=\pi/3$, and the tight-binding parameters $t=-3.548$~eV and $T=-4.657$~eV~\cite{al2014electronic}. Here $E_{i,f}$ and $\Psi_{i,f}$ denote the energy and wave functions of the initial and final states, respectively. The results are presented in the style adopted from Ref.~\cite{buchs2021metallic}, which studies another carbon-based system exhibiting THz transitions. Notably, the oscillator strength of THz transitions in odd-dimered polyynic rings linearly depend on the size of the applied electric field. It can also be seen from the matrix elements of velocity given above, that the oscillator strength of various transitions can be controlled by changing the polarization angle of the incident excitation relative to the applied electric field. This allows an additional (polarization) control of the inversion of population in the case of broadband optical excitation.

\begin{figure}
    \centering
    \includegraphics[width=0.5\textwidth]{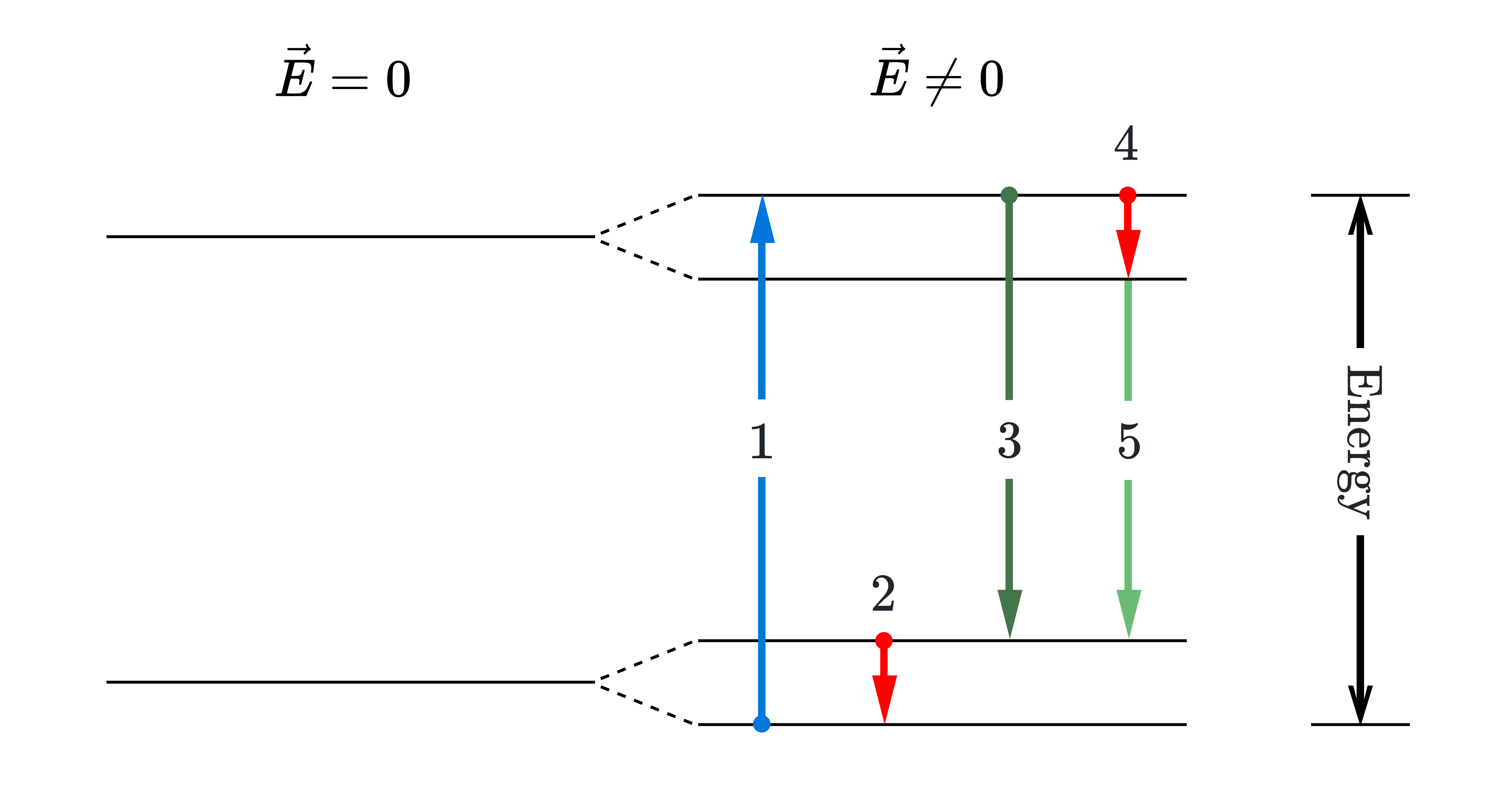}
    \caption{
A schematic illustration of a THz generation scheme using odd-dimered polyynic rings. A high-frequency optical excitation promotes an electron from the HOMO$-1$ to the LUMO$+1$ energy level, depicted by the blue arrow (1). An electron from the HOMO level relaxes down to the HOMO$-1$ level via the emission of a THz photon, depicted by the red arrow (2). The photo-excited electron in the LUMO$+1$ level may directly transition down to the HOMO level via the emission of a single optical photon (dark green arrow, 3), or via the emission of a THz photon (red arrow, 4) followed by an optical one (light green arrow, 5).}
    \label{fig:THz_Scheme}
\end{figure}

%\section*{THz generation scheme}
The presence of an optical HOMO-LUMO gap, coupled with the existence of energy levels spaced at THz frequencies (associated with the splitting of degenerate states of an ideal ring), makes odd-dimered polyynic rings subject to an in-plane electric field ideally suited for the basis of tunable THz emitters. One possible THz generation scheme involves the optical pumping of polyynic rings with linearly polarized light. As can be seen in Fig.~\ref{fig:THz_Scheme}, a high-energy optical photon can promote an electron from the HOMO$-1$ level to LUMO$+1$ level. This enables an electron to relax from the HOMO to the HOMO$-1$ via the emission of a THz photon. An additional THz photon may also be generated as the electron which was promoted to the LUMO$+1$ relaxes into the LUMO. 
%{\color{red} It should be noted that various relaxation pathways can be optimized by varying the polarisation angle of incident light. For example, the oscillator strength of the LUMO+1 to HOMO transition can be enhanced relative to the HOMO-1 to LUMO+1 transition. }

%\section*{Conclusion}
We have shown that for an ideal, odd-dimered polyynic ring, the symmetry-induced degeneracy of both HOMO and LUMO levels can be lifted by an external electric field. The induced doublet states support strong optical transitions between them, which fall within the highly desirable THz range. Our model can also describe a polyynic ring which has undergone spontaneous symmetry-breaking. In this instance the size of the THz gap induced by the Jahn-Teller effect can be mapped onto an effective dipole across an ideal ring. In both cases, the size of the gap is linearly dependent on the strength of the applied electric field. Therefore, the Stark effect in cyclocarbon could be used to create a THz source with the frequency controlled by the applied voltage. Appropriately arranged arrays of cyclocarbons placed within a carefully selected microcavity are thus promising candidates for active elements of amplifiers and generators of coherent THz radiation.

%{\color{red} Cyclocarbons exhibit a strong $\pi-\pi$ stacking effect~\cite{liu2021intermolecular}}

%Finally, our model can be used to explain that the striking contrast between the energy spectra of planar arrays comprised of odd and even $p$ (3$p$,0) zigzag nanotubes~\cite{polozkov2019carbon} arises due to the lack of an inversion center in the unit cell. %To our knowledge despite the large knowledge on curvature induced gaps the difference between odd and even have not been consider

%\section*{Acknowledgements}
This work was supported by the EU H2020 RISE projects TERASSE (H2020-823878) and DiSeTCom (H2020-823728). R.A.N. is funded by the DOST-SEI ASTHRDP program. R.R.H. acknowledges financial support from URCO (14 F 1TAY20-1TAY21). M.E.P. was supported by the NATO Science for Peace and Security project NATO.SPS.MYP.G5860.

The data used and analysed in this study are available from the corresponding author upon request.

\bibliography{ref}

\end{document}